\documentclass[twocolumn,showpacs,preprintnumbers,amsmath,amssymb,prb,aps]{revtex4-1}
\usepackage[dvips,colorlinks=true,bookmarks=false,citecolor=blue,urlcolor=blue]{hyperref} 

\usepackage{lmodern}

\usepackage[T1]{fontenc}
\usepackage[latin9]{inputenc}
\usepackage{esint}

\usepackage{amsmath,amssymb}
\usepackage{graphicx}
\usepackage{verbatim} 
\usepackage{color}

\usepackage[english]{babel}
\usepackage{bm}
\usepackage{natbib}
\usepackage{graphicx}
\usepackage{hyperref}

\def\dfrac{\displaystyle\frac}  

\usepackage{color}

\begin{document}

\title{Discrete solitons in graphene metamaterials }

\author{Yu.V. Bludov$^{1}$}\email{bludov@fisica.uminho.pt}
\author{D.A. Smirnova$^{2}$}
\author{Yu.S. Kivshar$^{2,3}$}
\author{N.M.R. Peres$^{1}$}
\author{M.I. Vasilevskiy$^{1}$}

\affiliation{$^{1}$Centro de Física and Departamento de Física, Universidade
do Minho, Campus de Gualtar, Braga 4710-057, Portugal \\
$^{2}$Nonlinear Physics Center, Research School of Physics and Engineering,
Australian National University, Canberra ACT 0200, Australia\\
$^{3}$ITMO University, St.~Petersburg 197101, Russia}


\begin{abstract}
We study nonlinear properties of multilayer metamaterials 
created by graphene sheets separated by dielectric layers.
We demonstrate that such structures can support localized nonlinear modes described by the discrete nonlinear Schr\"{o}dinger equation and that
its solutions are associated with stable discrete plasmon solitons.
We also analyze the nonlinear surface modes in truncated graphene metamaterials
being a nonlinear analog of surface Tamm states.
\end{abstract}

\maketitle

\section{Introduction}

Graphene is a unique two-dimensional (2D) material known to exhibit remarkable physical properties including a strong optical response related to its surface conductivity and dependence on graphene's chemical potential~\cite{Rev1,Rev2}.
At certain frequencies, doped graphene behaves like a metal, and it can support $p$-polarized surface plasmon polaritons
due to the coupling of the electromagnetic field to the electron excitations~\cite{Koppens_exp,Basovexp,RevGrigorenko,RevLuo}.

As has been shown recently, graphene is a strongly nonlinear material~\cite{Mikh_Nonlin,Mikhailov2008,Hendry2010,Hong2013X,Glazov2013,Sipe2014_NJPh,Sipe2014_OE}.
In particular, several nonlinear effects associated with a self-action correction to graphene's conductivity have been predicted recently~\cite{Nesterov2013,Coupler_PRB,PRL_SinglePhoton,Smirnova2013}. In order to increase the effective nonlinearity of photonic structures with graphene, a natural idea is to use graphene multilayers which, depending on different wavelength regimes, may possess the basic properties of photonic crystals and metamaterials~\cite{Wang2012,Bludov2013,our_meta}.

One of the remarkable general properties of nonlinear systems is their ability to support nonlinear localized modes -- self-trapped localized states or solitons which can propagate over long distances without changing their shape due to a balance between nonlinearity and dispersion (or diffraction). A special kind of soliton, the so-called \emph{discrete soliton}, appears as intrinsic localized mode in homogeneous periodic physical systems, such as nonlinear atomic chains~\cite{PhysRevLett.61.970,PhysRevLett.79.2510}, Bose-Einstein condensates loaded into optical lattices~\cite{PhysRevLett.86.2353,PhysRevE.66.046608}, arrays of nonlinear optical waveguides~\cite{Christodoulides1988}, and semiconductor-dielectric periodic nanostructures~\cite{Skryabin2009_PRA}. If compared to continuous localized waves, the discrete solitons possess a number of additional properties such as the Peierls-Nabarro barrier~\cite{PhysRevE.48.3077} and staggering transformation~\cite{Alfimov2004}. In plasmonics, discrete solitons were studied in metal-dielectric multilayer structures~\cite{Liu2007,Skryabin2010_OL,Kou2011,Kou2012,Kou2013}, arrays of nanowires~\cite{Ye2010,Ye2011,Silveirinha2013,Fernandes2014}, and arrays of nanoparticles~\cite{Noskov_prl,Noskov2012}.

\begin{figure}[!b]\centering
  \includegraphics[width=0.8\linewidth]{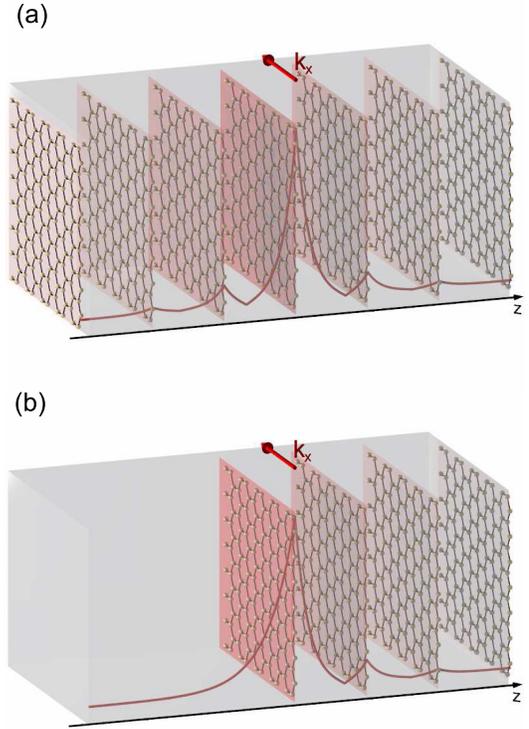}
  \caption{(Color online) Geometry of the problem. A multilayer structure is composed of graphene sheets separated by dielectric layers with permittivity $\varepsilon$ and thickness $d$.
Red curves show example profiles of the plasmonic solitons: (a) discrete solitons in an infinite structure, and (b) surface solitons in a truncated metamaterial. Shown is the absolute value of the tangential electric field component.} \label{fig:str}
\end{figure}

Less than a decade ago, an interesting type of discrete soliton, \emph{surface soliton}, was predicted theoretically~\cite{Makris2005} and then observed experimentally\cite{Suntsov2006,Rosberg2006}.
It is sustained by the boundary between a periodic structure and a uniform medium (although the maximum of a soliton can be either exactly at the interface~\cite{Makris2005} or at some distance from it~\cite{Molina2006};,
i.e. a surface soliton can be considered as a nonlinear analogue of surface Tamm states~\cite{Tamm1934,PhysRevB.89.245414}.

In this article, we study nonlinear graphene-based multilayer structures and demonstrate that, similar to metal-dielectric metamaterials, they can be described by the discrete nonlinear Schr\"{o}dinger (NLS) equation and support nonlinear localized modes in the form of discrete solitons [see Fig.~\ref{fig:str}(a)].
We also analyze such modes near the surfaces and predict the existence of nonlinear surface modes being a nonlinear analog of surface Tamm states, as shown schematically in Fig.~\ref{fig:str}(b).

This article is organized as follows. In Sec.~\ref{sec:current}, we discuss the nonlinear response of graphene to an external harmonic electric field. In Sec.~\ref{sec:solitons}, we derive the discrete NLS equation and describe the properties of discrete solitons. Section~\ref{sec:surface-solitons} is devoted to the study of surface solitons localized in the vicinity of a terminated layer of the graphene metamaterial.

\section{Nonlinear current in graphene\label{sec:current}}

For the sake of completeness and clarity, first we derive a Kerr-type nonlinear correction to the graphene conductivity,
considered earlier in Refs.~\cite{Mikh_Nonlin,Mikhailov2008} for the ballistic regime.

We consider a 2D doped graphene monolayer, placed parallel to the plane
$xy$. Also we admit that a time-dependent external electric field is
applied to graphene. For definiteness, the electric field is supposed to be directed
along the $x$ axis, i.e., $\vec{E}=[E(t),0,0]$. In principle, the temporal dependence of $E(t)$ can have an arbitrary
form, although in the calculations below it is considered to be of the form $E(t)=E_{0}\exp(-i\omega t)+\mathrm{c.c.}$, where $E_{0}$ and $\omega$
are the amplitude and the frequency.

In the classical frequency range, $\hbar \omega \leq {E}_{F}$, in the relaxation time approximation, graphene charge-carriers transport
properties are governed by the Boltzmann kinetic equation written for the electrons:
\begin{equation}
\frac{\partial f(\vec{k},t)}{\partial t}-\frac{e}{\hbar}\overrightarrow{E}\frac{\partial f(\vec{k},t)}{\partial\overrightarrow{k}}=-\gamma\left[f(\vec{k},t)-f_{0}(\vec{k})\right]\,.\label{eq:BE}
\end{equation}
 where $f(\vec{k},t)$ is the nonequilibrium distribution function,
$f_{0}(\vec{k})$ is the equilibrium Fermi-Dirac distribution function,
and $\gamma$ is the inverse relaxation time. Equation (\ref{eq:BE})
can be solved analytically, and its \emph{exact} solution at $t \gg 1/\gamma$ is given by~\cite{Glazov2013,Peres2014_arxiv}
\begin{equation}
f(\vec{k},t)=\gamma e^{-\gamma t}\int_{-\infty}^{t}dt^{\prime}e^{\gamma t^{\prime}}f_{0}[k_{x}+H\left(t,t^{\prime}\right),k_{y})]\,,\label{eq:BeqSol}
\end{equation}
where
\begin{eqnarray*}
 & H(t,t^{\prime})=\frac{e}{\hbar}\int_{t'}^{t}E(t^{\prime\prime})dt^{\prime\prime}=-\frac{e}{i\hbar\omega}\left[E_{0}\exp(-i\omega t)\right.\\
 & -\left.\overline{E_{0}}\exp(i\omega t)-E_{0}\exp(-i\omega t^{\prime})+\overline{E_{0}}\exp(i\omega t^{\prime})\right]
\end{eqnarray*}
and the overbars stand for complex conjugation.

The induced 2D current in graphene is expressed through the function
$f(\vec{k},t)$ as
\begin{equation}
\vec{j}=-4\frac{e}{(2\pi)^{2}}\int d\vec{k}\, f(\vec{k},t)\,\frac{\partial\epsilon\left(\vec{k}\right)}{\hbar\partial\overrightarrow{k}}\,,\label{eq:semi-classical_current}
\end{equation}
where $\epsilon\left(\vec{k}\right)=v_{F}\hbar\sqrt{k_{x}^{2}+k_{y}^{2}}$
is the Dirac cone spectrum of charge carriers in graphene, $v_{F}$
is the Fermi velocity, and the factor 4 
is due to the spin and valley degeneracy. Even though for large wave vectors the energy spectrum becomes anisotropic (leading to the trigonal warping of constant energy surfaces), for the levels of graphene doping, what nowadays are experimentally achievable, the Dirac cone approximation gives reasonable accuracy in the calculation of graphene nonlinear conductivity. A comparison between the Dirac cone and trigonal warping approximations is presented in the Appendix. For degenerate electrons at zero temperature (and,
consequently, the steplike Fermi-Dirac distribution function $f_{0}(\vec{k})=\Theta\left[E_{F}-\epsilon\left(\vec{k}\right)\right]$),
we obtain
\begin{equation}
j_{x}=-\frac{ev_{F}}{\pi{}^{2}}\gamma e^{-\gamma t}\int_{-\infty}^{t}dt^{\prime}e^{\gamma t^{\prime}}I\left(t,t^{\prime}\right)\,,\label{eq:current_x}
\end{equation}
where
\begin{eqnarray}
 & I\left(t,t^{\prime}\right)=\int d\vec{k}\,\frac{k_{x}}{\sqrt{k_{x}^{2}+k_{y}^{2}}}\nonumber \\
 & \times\Theta\left[E_{F}-v_{F}\hbar\sqrt{\left\{ k_{x}+H\left(t,t^{\prime}\right)\right\} ^{2}+k_{y}^{2}}\right]\label{eq:Integral}\\
 & =\int_{0}^{k_{F}}k^{\prime}\, dk^{\prime}\int_{0}^{2\pi}d\varphi\frac{k^{\prime}\cos\varphi-H\left(t,t^{\prime}\right)}{\sqrt{\left\{ k^{\prime}\cos\varphi-H\left(t,t^{\prime}\right)\right\} ^{2}+k^{\prime2}\sin^{2}\varphi}},\nonumber
\end{eqnarray}
$E_{F}=v_{F}\hbar k_{F}$ is the Fermi energy ($k_{F}$ is the Fermi
wave vector), $\Theta\left(x\right)$ is the Heaviside function, and the
changes of variables are $k_{x}=k^{\prime}\cos\varphi-H\left(t,t^{\prime}\right)$ and
$k_{y}=k^{\prime}\sin\varphi$.

After integration with respect to
$k^{\prime}$, expression~(\ref{eq:Integral}) can be presented in the form
\begin{widetext}
\begin{eqnarray*}
 & I\left(t,t^{\prime}\right)=\int_{0}^{2\pi}\left\{ \sqrt{k_{F}^{2}-2H\left(t,t^{\prime}\right)k_{F}+H^{2}\left(t,t^{\prime}\right)}\left[\frac{k_{F}\cos\varphi}{2}-H\left(t,t^{\prime}\right)\left(1-\frac{3}{2}\cos^{2}\varphi\right)\right]+H^{2}\left(t,t^{\prime}\right)\left(1-\frac{3}{2}\cos^{2}\varphi\right)\right.\\
 & +\left.\frac{3}{2}H^{2}\left(t,t^{\prime}\right)\left(\cos^{3}\varphi-\cos\varphi\right)\ln\left[\frac{\sqrt{k_{F}^{2}-2H\left(t,t^{\prime}\right)k_{F}+H^{2}\left(t,t^{\prime}\right)}+k_{F}-H\left(t,t^{\prime}\right)\cos\varphi}{H\left(t,t^{\prime}\right)\left(1-\cos\varphi\right)}\right]\right\} d\varphi .
\end{eqnarray*}
\end{widetext}
After the expansion with respect to $H\left(t,t^{\prime}\right)$
(up to the third order), the integral (\ref{eq:Integral}) is reduced to
\begin{equation}
I\left(t,t^{\prime}\right)=-k_{F}\pi H\left(t,t^{\prime}\right)+\frac{\pi}{8k_{F}}H^{3}\left(t,t^{\prime}\right).\label{eq:I-expand}
\end{equation}
Finally, substituting Eq.~(\ref{eq:I-expand}) into Eq.~(\ref{eq:current_x})
and performing integration, we obtain
\begin{eqnarray}
 & j_{x}=\sigma_{0}\frac{4E_{F}}{\pi\hbar}\frac{E_{0}\exp(-i\omega t)}{\gamma-i\omega}\label{eq:current_final}\\
 & -\sigma_{0}\frac{9e^{2}v_{F}^{2}}{\pi E_{F}\hbar}\frac{\left|E_{0}\right|^{2}E_{0}\exp(-i\omega t)}{\left(\gamma-2i\omega\right)\left(\gamma^{2}+\omega^{2}\right)}+\mathrm{c.c.},\nonumber
\end{eqnarray}
where $\sigma_{0}=e^{2}/4\hbar$ is the conductivity quantum. Note that in Eq.~(\ref{eq:current_final}) we write out only the terms with the time dependence $\sim\exp\left(\pm i\omega t\right)$, while the terms
corresponding to the third harmonic are omitted.

In the limit $\omega / \gamma \gg 1$,
Eq.~(\ref{eq:current_final}) can be written as
\begin{equation}
j_{x}=i\left[\nu^{(1)}-\nu^{(3)}\left|E_{0}\right|^{2}\right]E_{0}\exp(-i\omega t),
\label{eq:current_short}
\end{equation}
where
\[
\nu^{(1)}=\sigma_{0}\frac{4E_{F}}{\pi\hbar\omega},\qquad\nu^{(3)}=\sigma_{0}\frac{9e^{2}v_{F}^{2}}{2\pi E_{F}\hbar\omega^{3}}.
\]
Below, we use this result, obtained as seen by free-space light normally incident on a graphene layer,
for the effective nonlinear conductivity of surface plasmons propagating along graphene layers, assuming the additional correction
due to the in-plane wavevector $k_x$ to be small, which is well-justified if $c k_x/\omega \ll 300$.

\section{Discrete solitons\label{sec:solitons}}

Now we consider a periodic multilayer graphene stack, consisting of
an infinite number of parallel graphene layers arranged at equal
distances $d$ from each other at the planes $z=md$ with $m=\left(-\infty,\infty\right)$, inside a dielectric
medium with relative permittivity $\varepsilon$. In this case, the electric $\vec{E}$
and magnetic $\vec{H}$ fields are governed by Maxwell's equations:
\begin{eqnarray*}
 & \mathrm{rot}\vec{E}=i\omega\mu_{0}\vec{H},\qquad\mathrm{div}\vec{E}=\frac{\rho}{\varepsilon\varepsilon_{0}},\\
 & \mathrm{rot}\vec{H}=-i\omega\varepsilon_{0}\varepsilon\vec{E}+\vec{J},\qquad\mathrm{div}\vec{H}=0,
\end{eqnarray*}
where $\varepsilon_{0}$ and $\mu_{0}$ are free-space permittivity and
permeability, and $\vec{J}$ and $\rho$ are full three-dimensional (3D) current and charge
densities, respectively, given by
\begin{equation}
\vec{J}=\sum_{m=-\infty}^{\infty}\vec{j}^{(m)}\delta(z-md),\qquad\rho=\sum_{m=-\infty}^{\infty}\varrho^{(m)}\delta(z-md),\label{eq:J-rho}
\end{equation}
where $\vec{j}^{(m)}$ and $\varrho^{(m)}$ are 2D current and charge
densities in the $m$th graphene layer. In all the above equations
the time-dependence $\exp(-i\omega t)$ is implied.

The electric and magnetic fields can be expressed through scalar $\varphi$
and vector $\vec{A}$ potentials as
\begin{equation}
\vec{E}=-\mathrm{grad}\varphi+i\omega\vec{A},\qquad\vec{H}=\frac{\mathrm{rot}\vec{A}}{\mu_{0}}.
\end{equation}
These relations, jointly with the Lorentz gauge
\begin{equation}
\mathrm{div}\vec{A}-(i\omega\varepsilon/c^{2})\varphi=0\label{eq:Lor-gauge},
\end{equation}
result in inhomogeneous Helmholtz equations for both scalar and vector potentials:
\begin{eqnarray}
 & \Delta\varphi+\frac{\omega^{2}\varepsilon}{c^{2}}\varphi=-\frac{\rho}{\varepsilon\varepsilon_{0}},\\
 & \Delta\vec{A}+\frac{\omega^{2}\varepsilon}{c^{2}}\vec{A}=-\mu_{0}\vec{J}.\label{eq:A}
\end{eqnarray}

We assume the electromagnetic field to be uniform
along the $y$ direction, $\partial/\partial y\equiv0$, and propagating in the $x$ direction, $\vec{A},\,\vec{J}\,\rho,\,\varphi\sim\exp\left(ik_{x}x\right)$.
Under these assumptions, Eq.~(\ref{eq:A}) can be solved by using a standard Green's function formalism. Accordingly, a general solution of Eq.~(\ref{eq:A}) has the form
\begin{equation}
A_x\left(z\right)=-\mu_{0}\int_{-\infty}^{\infty}dz^{\prime}G\left(z-z^{\prime}\right)J_x,\label{eq:A-sol-gen}
\end{equation}
where
\[
G(z)=-\frac{\exp\left(-p\left|z\right|\right)}{2p},\quad p=\sqrt{k_{x}^{2}-\frac{\omega^{2}\varepsilon}{c^{2}}}
\]
is the one-dimensional Green function. The latter is a solution of the equation
\[
\left(\frac{d^{2}}{dz^{2}}-p^{2}\right)G\left(z\right)=\delta\left(z\right)
\]
with the boundary conditions $G\left(\pm\infty\right)=0$, denoting the evanescent character of
waves (when $k_{x}^{2}>\left(\omega/c\right)^{2}\varepsilon$ and $\mathrm{Re}\left(p\right)>0$)
or absence of waves coming from $z=\pm\infty$ for traveling waves, when $k_{x}^{2}<\left(\omega/c\right)^{2}\varepsilon$ and $\mathrm{Im}\left(p\right)<0$.
Substituting Eq.~(\ref{eq:J-rho}) into Eq.~(\ref{eq:A-sol-gen}) and using
the properties of Delta functions, we obtain
\[
A_x\left(z\right)=\frac{\mu_{0}}{2p}\sum_{m=-\infty}^{\infty}j^{(m)}_x\exp(-p\left|z-md\right|).
\]
Due to the 2D nature
of currents in graphene layers $A_{z}\equiv0$, while the vector potential components $A_{x}$ and $A_{y}$
describe $p$- and $s$-polarized waves, correspondingly. Further we concentrate on the $p$-polarized waves only. Thus, using the Lorentz
gauge (\ref{eq:Lor-gauge}), we can express the $x$ component of the electric
field through $A_{x}$ as
\begin{equation}
E_{x}\left(z\right)=\frac{c^{2}p^{2}}{i\omega\varepsilon}A_{x}\left(z\right).\label{eq:Ex-Ax}
\end{equation}
After substituting this relation into Eq.~(\ref{eq:current_short}),
$A_{x}$ can be represented in the form
\begin{equation} \label{eq:Az-sol-inf}
\begin{aligned}
A_{x}\left(z\right)=\frac{p}{2\omega\varepsilon_{0}\varepsilon}\sum_{m=-\infty}^{\infty}\left[\nu^{(1)}-\nu^{(3)}\frac{c^{4}p^{4}}{\omega^{2}\varepsilon^{2}}\left|A_{x}\left(md\right)\right|^{2}\right] \\
\times A_{x}\left(md\right)\exp(-p\left|z-md\right|).
\end{aligned}
\end{equation}
Alternatively, Eq.~(\ref{eq:Az-sol-inf}) can be rewritten in the
form of the stationary discrete NLS equation
\begin{equation} \label{eq:DNLS}
\begin{aligned}
 & A_{x}\left(\left[n+1\right]d\right)+A_{x}\left(\left[n-1\right]d\right)-2A_{x}\left(nd\right)\cosh(pd) \\
 &  =-\frac{p}{\omega\varepsilon_{0}\varepsilon}\left[\nu^{(1)}-\nu^{(3)}\frac{c^{4}p^{4}}{\omega^{2}\varepsilon^{2}}\left|A_{x}\left(nd\right)\right|^{2}\right] A_{x}\left(nd\right)\sinh\left(pd\right)
\end{aligned}
\end{equation}
for $n\in(-\infty,\infty)$.

\begin{figure}
\center{\includegraphics[width=8.5cm]{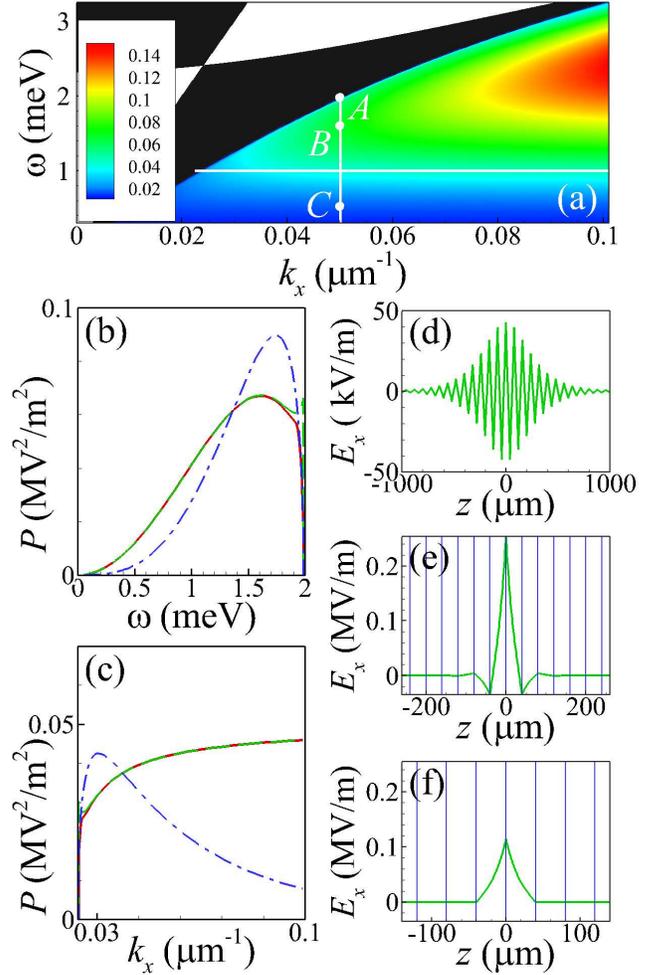}}
\caption{(Color online) (a)--(c) Dependence of soliton norm $P$ (in $\mathrm{MV}^{2}/\mathrm{m}^{2}$)
upon frequency $\omega$ and wave vector $k_{x}$ {[}panel (a){]},
upon frequency $\omega$ for fixed value $k_{x}=0.05\,\mu\mathrm{m}^{-1}${[}panel
(b){]}, or upon frequency $k_{x}$ for fixed value $\omega=1\,$meV
{[}panel (c){]}. Dependencies in panels (b) and (c) are taken along
the vertical and horizontal lines in panel (a), respectively. Dependencies
in panel (a) as well as those in panels (b) and (c) {[}depicted by
solid lines{]} are calculated by the numerical solution of Eq.~(\ref{eq:DNLS}),
while continuum (dash-and-dot lines) and anticontinuum (dashed lines)
limit approximations in panels (b) and (c) are calculated according
to Eqs.~(\ref{eq:P-cont}) and (\ref{eq:P-anticont}). (d)--(f) Soliton
spatial profiles for $k_{x}=0.05\,\mu\mathrm{m}^{-1}$ and $\omega=1.98\,$meV
{[}panel (d){]}, $\omega=1.6\,$meV {[}panel (e){]}, or $\omega=0.52\,$meV
{[}panel (f){]}. The parameters of panels (d), (e), and (f) correspond
to points $A$, $B$, and $C$ in panel (a), respectively. Other parameters
are $E_{F}=0.157\,$eV, $d=40\,\mu$m, and $\varepsilon=3.9$.}
\label{fig:br_soliton}
\end{figure}

\begin{figure}
\includegraphics[width=8.5cm]{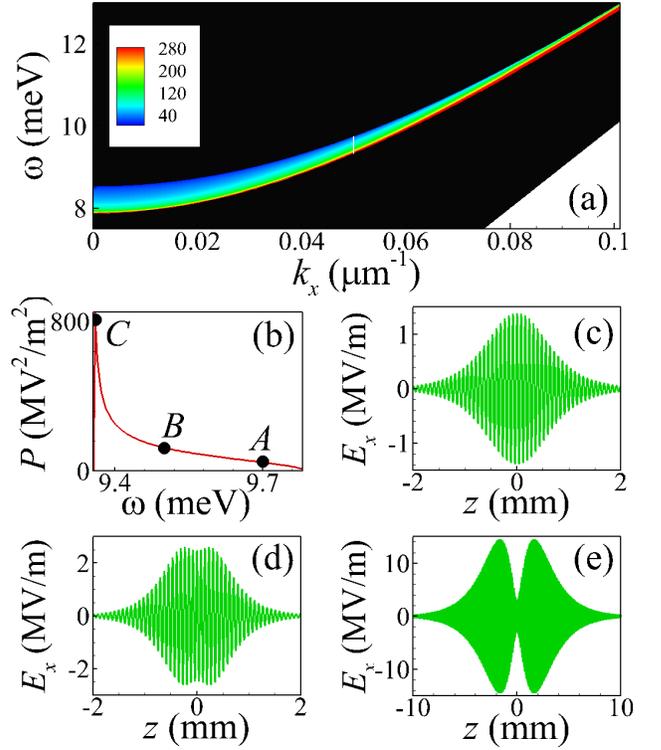} 
\protect\caption{(Color online) Dependence of soliton norm $P$ (in $\mathrm{MV}^{2}/\mathrm{m}^{2}$)
upon frequency $\omega$ (in the second gap) and wave vector $k_{x}$
{[}panel (a){]} or upon frequency $\omega$ for fixed value $k_{x}=0.05\,\mu\mathrm{m}^{-1}${[}panel
(b){]}. Dependence in panel (b) is taken along the vertical line
in panel (a). (c)--(e) Soliton spatial profiles for $k_{x}=0.05\,\mu\mathrm{m}^{-1}$
and $\omega=9.7\,$meV {[}panel (c){]}, $\omega=9.5\,$meV {[}panel
(d){]}, or $\omega=9.36\,$meV {[}panel (e){]}. The parameters of
panels (c), (d), and (e) correspond to points $A$, $B$, and $C$ in panel (b),
respectively. Other parameters are the same as those in Fig.~\ref{fig:br_soliton}.\label{fig:br_sol_sg}}
\end{figure}

The linear counterpart (when $\nu^{(3)}=0$) of the discrete NLS equation, Eq.(\ref{eq:DNLS}), defines the linear spectrum. Domains of allowed
frequencies (where in the linear case the wave propagation is possible)
are parametrized by the real Bloch wave vector $q$ {[}such that $A_{x}\left(nd\right)=A_{x}\left(0\right)\exp\left(iqnd\right)${]}.
As a result, the equation
\begin{equation}
\cos\left(qd\right)=\cosh(pd)-\frac{p}{2\omega\varepsilon_{0}\varepsilon}\nu^{(1)}\sinh\left(pd\right)\label{eq:lin-spec}
\end{equation}
determines the propagating bands of the spectrum $\omega=\Omega_{l}\left(k_{x},q\right)$
($l\ge1$ is the band index), which are depicted in Figs.~\ref{fig:br_soliton}(a)
and \ref{fig:br_sol_sg}(a) in black (see, e.g., Ref.~\cite{Bludov2013}).
%
%

Although generally the nonlinear equation, Eq.~(\ref{eq:DNLS}), possesses an infinite number
of solutions~\cite{Alfimov2004}, here we concentrate on the properties of the fundamental bright
solitons, bifurcating from the edge of the allowed band of the spectrum.
To describe the solitons' properties, we introduce a soliton norm
as
\[
P=\sum_{m=-\infty}^{\infty}\left|E_{x}\left(md\right)\right|^{2}=\frac{c^{4}p^{4}}{\omega^{2}\varepsilon^{2}}\sum_{m=-\infty}^{\infty}\left|A_{x}\left(md\right)\right|^{2}.
\]
The fundamental mode of the discrete soliton is depicted in Fig.~\ref{fig:br_soliton}.
Due to effectively defocusing nonlinearity (positive cubic term) in Eq.~(\ref{eq:DNLS}), bright
solitons {[}see Fig.~\ref{fig:br_soliton}(a){]} bifurcate from the
low-frequency boundary of the first band $\Omega_{1}(k_{x},q)$ (which
corresponds to the phase shift $qd=\pi$ between oscillations in adjacent
graphene layers) and exist in the semi-infinite gap $\omega\le\Omega_{1}(k_{x},\pi/d)$.
Since in this region $k_{x}>\omega\varepsilon^{1/2}/c$, this type
of soliton is characterized by the evanescent waves in the dielectric
between the graphene layers, and these solitons are further referred to
as \emph{plasmonic solitons}. For fixed $k_{x}$ {[}Fig.~\ref{fig:br_soliton}(b){]}
the soliton norm $P$, being zero at the band edge $\omega=\Omega_{1}(k_{x},\pi/d)$,
initially grows up to values $\sim10^{11}\,\mathrm{V}^{2}/\mathrm{m}^{2}$,
but after that decreases and attains zero at zero frequency. At the
same time, the frequency defines the degree of soliton localization,
as follows from the comparison of Figs.~\ref{fig:br_soliton}(d)--\ref{fig:br_soliton}(f).
Thus, in the vicinity of the band edge $\Omega_{1}(k_{x},q)$ the soliton
is delocalized -- its electric field is distributed over a large number
of graphene layers {[}Fig.~\ref{fig:br_soliton}(d){]}. When frequency
is gradually detuned from the band edge, the soliton becomes more localized
-- its electric field is either distributed over a few graphene layers
{[}Fig.~\ref{fig:br_soliton}(e){]} or effectively concentrated in
the vicinity of one graphene layer, as shown in Fig.~\ref{fig:br_soliton}(f).
It should be underlined that the soliton inherits the properties of a Bloch
wave at the band edge from which it bifurcates: signs of the electric
field tangential components at adjacent graphene layers are opposite (staggered soliton).
For fixed frequency $\omega$ {[}Fig.~\ref{fig:br_soliton}(c){]}
the soliton norm increases monotonically with increasing $k_{x}$.

Equation~(\ref{eq:DNLS}) possesses two approximate types of solutions.
The first type, the so-called \emph{continuum limit}, is valid
for low amplitude solutions. To obtain this solution,
we use the ansatz $A_{x}\left(nd\right)=\epsilon\left(-1\right)^{n}\psi(\zeta)$,
with $\epsilon$ being a small parameter and $\zeta=\epsilon n$.
As a result, the function $\psi\left(\zeta\right)$ satisfies the
nonlinear Schrodinger equation
\begin{equation} \label{eq:psi}
\dfrac{d^{2}\psi}{dx^{2}}+\nu^{(3)}\frac{c^{4}p^{5}}{\omega^{3}\varepsilon_{0}\varepsilon^{3}}\sinh\left(pd\right)\psi^{3}(x)=\frac{2\cosh(\beta)-2}{\epsilon^{2}}\psi(x),
\end{equation}
which is parametrized by parameter $\beta$ such that
\begin{equation}
\cosh\left(\beta\right)=\frac{p}{2\omega\varepsilon_{0}\varepsilon}\nu^{(1)}\sinh\left(pd\right)-\cosh(pd).\label{eq:beta}
\end{equation}
The parameter $\beta$ can be formally considered as the imaginary part of the Bloch wavevector $q=\left(\pi+i\beta\right)/d$ (note, inside the gap the Bloch wave vector is complex that in the linear case corresponds to the evanescent wave). Using the exact solution of Eq.~(\ref{eq:psi}), we now approximate solution of Eq.~(\ref{eq:DNLS})
in the continuum limit:
\[
A_{x}\left(nd\right)=\sqrt{\frac{2\,\omega^{3}\varepsilon_{0}\varepsilon^{3}}{c^{4}p^{5}\nu^{(3)}\sinh\left(pd\right)}}\frac{\left(-1\right)^{n}\sqrt{2\cosh(\beta)-2}}{\cosh\left(\sqrt{2\cosh(\beta)-2}n\right)}.
\]
Consequently, the soliton norm in the continuum limit can be expressed
as
\begin{equation}
\begin{aligned}
P=\frac{2\,\omega\varepsilon_{0}\varepsilon\left[2\cosh(\beta)-2\right]}{p\nu^{(3)}\sinh\left(pd\right)}
& \sum_{n=-\infty}^{\infty}\frac{1}{\cosh^{2}\left(\sqrt{2\cosh(\beta)-2}n\right)}\\
&\approx\frac{4\,\omega\varepsilon_{0}\varepsilon\sqrt{2\cosh(\beta)-2}}{p\nu^{(3)}\sinh\left(pd\right)}.
\end{aligned}
\label{eq:P-cont}
\end{equation}
In the last equation the summation has been replaced by the integration.
As seen from Fig.~\ref{fig:br_soliton}, the continuum approximation (depicted by blue dash-and-dot line)
is valid in the narrow domain in the vicinity of band edge $\Omega_{1}(k_{x},\pi/d)$
{[}more specifically, in domains 1.95 meV$\lesssim\omega\lesssim$1.987
meV in Fig.~\ref{fig:br_soliton}(b) and 0.0236$\,\mu\mathrm{m^{-1}}\lesssim k_{x}\lesssim$0.0245$\,\mu\mathrm{m^{-1}}$
in Fig.~\ref{fig:br_soliton}(c){]}.

The other type of approximate solutions, so-called \emph{anticontinuum limit},
is valid far from the band edge $\Omega_{1}(k_{x},\pi/d)$ (deeply
in the gap). Hence, introducing scaled dimensionless variables
\[
a_{n}=\left({\nu^{(3)}\frac{c^{4}p^{5}}{\varepsilon_{0}\omega^{3}\varepsilon^{3}}\frac{\sinh\left(pd\right)}{2\cosh(\beta)}}\right)^{1/2}A_{x}\left(nd\right),
\]
and taking into account Eq. (\ref{eq:beta}), we obtain
\[
\frac{a_{n+1}+a_{n-1}}{2\cosh(\beta)}+a_{n}-a_{n}^{3}=0.
\]
As a result, when $\beta\to\infty$, $a_{n}$ become independent
and acquire one of the following three values: $a_{n}=-1$, $a_{n}=0$, or $a_{n}=+1$.
In this limit, the fundamental mode {[}see, e.g., Fig.~\ref{fig:br_soliton}(e){]}
corresponds to the case where $a_{n}=\delta_{n,0}$. This case allows
for the approximate analytical continuation valid for large values of $\beta$:
\begin{eqnarray*}
a_{0}=1-\frac{1}{4\cosh^{2}(\beta)},\\
a_{1}=a_{-1}=-\frac{1}{2\cosh(\beta)}-\frac{1}{8\cosh^{3}(\beta)},\\
a_{2}=a_{-2}=\frac{1}{4\cosh^{2}(\beta)}.
\end{eqnarray*}
As a result, the soliton
norm can be represented in the form
\begin{eqnarray}
P=\frac{\varepsilon_{0}\omega\varepsilon}{p\nu^{(3)}}\frac{2\cosh(\beta)}{\sinh\left(pd\right)}\left[a_{0}^{2}+2a_{1}^{2}+2a_{2}^{2}\right]\label{eq:P-anticont}\\
=\frac{\varepsilon_{0}\omega\varepsilon}{p\nu^{(3)}}\frac{1}{\sinh\left(pd\right)}\left[2\cosh(\beta)+\frac{7}{8\cosh^{3}(\beta)}\right].\nonumber
\end{eqnarray}
As seen from Fig.~\ref{fig:br_soliton}, the anticontinuum limit approximation (depicted by the green dashed line)
well describes the solution in domains 0$\lesssim\omega\lesssim$1.95
meV [in Fig.~\ref{fig:br_soliton}(b)] and $k_{x}\gtrsim$0.0245$\,\mu\mathrm{m^{-1}}$
[in Fig.~\ref{fig:br_soliton}(c)].

Solitons can also exist in the upper (finite) gaps of the spectrum.
Notice that in those gaps $k_{x}<\omega\varepsilon^{1/2}/c$, and
solitons are characterized by propagating waves in the dielectric
between graphene layers (this type of soliton is further referred to
as a \emph{photonic soliton}). An example of photonic solitons is shown
in Fig.~\ref{fig:br_sol_sg}. Photonic solitons are characterized
by considerably larger soliton norms $P$ if compared to the plasmonic
ones {[}soliton norm is of the order of 500 $\,\mathrm{MV^{2}/m^{2}}$ in
Fig.~\ref{fig:br_sol_sg}(a) and 0.1 $\,\mathrm{MV^{2}/m^{2}}$ in Fig.~\ref{fig:br_soliton}(a){]}.
Photonic solitons bifurcate from the upper edge of the gap -- the
soliton norm, being zero at the high-frequency boundary of the gap
$\Omega_{3}\left(k_{x},\pi/d\right)$, is increased when the frequency
is decreased {[}see Fig.~\ref{fig:br_sol_sg}(b){]}. The decrease of the
frequency also leads to the growth of the soliton amplitude {[}compare
Figs.~\ref{fig:br_sol_sg}(c)--\ref{fig:br_sol_sg}(e){]}.
At the same time, photonic solitons are considerably wider than plasmonic ones,
and at large amplitudes they become two-hump
{[}Figs.~\ref{fig:br_sol_sg}(d) and \ref{fig:br_sol_sg}(e){]}.
This happens due to the fact that, by contrast to plasmonic solitons, for photonic solitons local
maxima and minima of the electromagnetic field are generally not located
at graphene layers.

\section{Discrete surface solitons\label{sec:surface-solitons}}
\begin{figure}
\includegraphics[width=8.5cm]{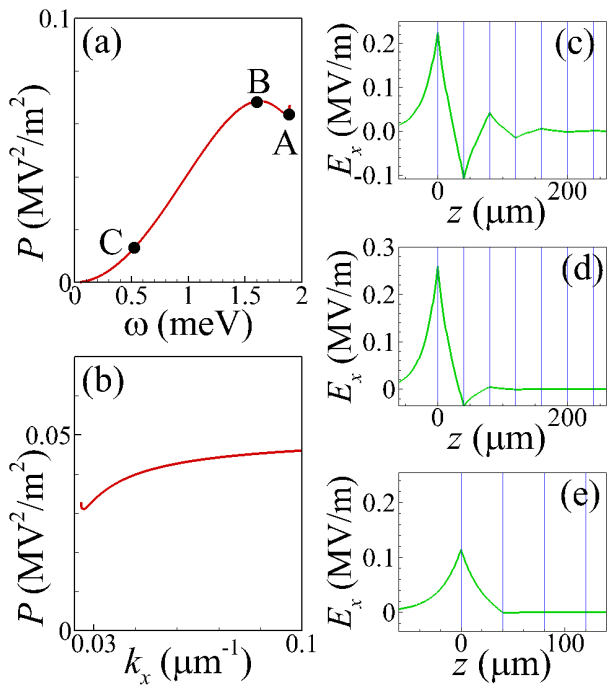} 
\protect\caption{(Color online) Dependence of surface soliton norm $P$ (in $\mathrm{MV}^{2}/\mathrm{m}^{2}$)
upon frequency $\omega$ for the fixed value $k_{x}=0.05\,\mu\mathrm{m}^{-1}${[}panel
(a){]} or upon frequency $k_{x}$ for the fixed value $\omega=1\,$meV
{[}panel (b){]}. (c)--(e) Soliton spatial profiles for $k_{x}=0.05\,\mu\mathrm{m}^{-1}$
and $\omega=1.89\,$meV {[}panel (c){]}, $\omega=1.6\,$meV {[}panel
(d){]}, or $\omega=0.52\,$meV {[}panel (e){]}. The parameters of
panels (c), (d), and (e) correspond to points $A$, $B$, and $C$ in panel (a),
respectively.\label{fig:surf-sol}}
\end{figure}
Finally, we consider a semi-infinite array of graphene layers, arranged
at equal distances $d$ from each other at planes $z=md$ with $m=\left[0,\infty\right)$, as shown in Fig.~\ref{fig:str}(b).
In other words, graphene layers are embedded inside a semi-infinite
dielectric medium at $z\geq 0$, while at $z<0$ there is just a homogeneous dielectric. The 3D current and charge density for this semi-infinite array
can be written as
\begin{equation}
\vec{J}=\sum_{m=0}^{\infty}\vec{j}^{(m)}\delta(z-md),\qquad\rho=\sum_{m=0}^{\infty}\varrho^{(m)}\delta(z-md),\label{eq:J-rho-semiinf}
\end{equation}
and the solution of the wave equation, Eq. (\ref{eq:A}), has {[}in full
analogy with Eq.~(\ref{eq:Az-sol-inf}){]} the form
\begin{equation} \label{eq:Az-sol-semiinf}
\begin{aligned}
A_{x}\left(z\right)=\frac{p}{2\omega\varepsilon_{0}\varepsilon}&\sum_{m=0}^{\infty}\left[\nu^{(1)}-\nu^{(3)}\frac{c^{4}p^{4}}{\omega^{2}\varepsilon^{2}}\left|A_{x}\left(md\right)\right|^{2}\right]\\
&\times A_{x}\left(md\right)\exp(-p\left|z-md\right|),
\end{aligned}
\end{equation}
or
\begin{eqnarray*}
 &  & A_{x}\left(\left[n+1\right]d\right)+A_{x}\left(\left[n-1\right]d\right)-2A_{x}\left(nd\right)\cosh(pd)\\
 &  & =-\frac{p}{\omega\varepsilon_{0}\varepsilon}\left[\nu^{(1)}-\nu^{(3)}\frac{c^{4}p^{4}}{\omega^{2}\varepsilon^{2}}\left|A_{x}\left(nd\right)\right|^{2}\right]\\
 &  & \times A_{x}\left(nd\right)\sinh\left(pd\right),\qquad {\rm for}\,\,\, n>0;\\
 &  & A_{x}\left(d\right)-A_{x}\left(0\right)\exp(pd)\\
 &  & =-\frac{p}{\omega\varepsilon_{0}\varepsilon}\left[\nu^{(1)}-\nu^{(3)}\frac{c^{4}p^{4}}{\omega^{2}\varepsilon^{2}}\left|A_{x}\left(0\right)\right|^{2}\right]\\
 &  &\times A_{x}\left(0\right)\sinh\left(pd\right).
\end{eqnarray*}

Properties of plasmonic surface solitons are summarized in Fig.~\ref{fig:surf-sol}.
The principal difference between the cases of surface and
bulk solitons is the nonexistence of the low-amplitude surface soliton
in the vicinity of the band edge $\Omega_{1}\left(k_{x},\pi/d\right)$
{[}compare, e.g., Figs.~\ref{fig:surf-sol}(a) and \ref{fig:br_soliton}(b),
as well as Figs.~\ref{fig:surf-sol}(b) and \ref{fig:br_soliton}(c){]}.
More specifically, there exists an end point of the spectrum, at which
the fundamental mode bifurcates with the other type of the surface
soliton mode (for details see, e.g., Ref.~\cite{Bludov2007}). In
the vicinity of the end point of the spectrum soliton norm $P$ achieves
a local minimum. At the same time, from the comparison of Figs.~\ref{fig:surf-sol}(c)-\ref{fig:surf-sol}(e)
it follows that, similar to the case of bulk solitons, lower frequencies
correspond to more localized solitons (when the power is mostly
concentrated at the graphene layer, truncating the photonic crystal).

It is also worth noting that the principal difference between linear
and nonlinear cases is the possibility to have the nonlinear surface
state (namely, surface soliton) in the uniform structure (semi-infinite
array of \emph{equally} doped graphene layers, placed at \emph{equal}
distances from each other, and embedded into the \emph{uniform} dielectric
medium), while in the linear case the existence of the surface state is
possible only in the nonuniform structure -- it is necessary to have
either the defect of the periodicity at the surface of the photonic crystal
\cite{PhysRevB.89.245414} or the defect of graphene doping at the surface,
or to truncate the photonic crystal with the dielectric, characterized
by the dielectric constant, different from that of the medium inside
the photonic crystal.

\section{Conclusions}

We have analyzed nonlinear graphene-based multilayer metamaterials and demonstrated that they can support spatially localized nonlinear modes in the form of discrete plasmon solitons. We have described the properties of this novel class of discrete solitons, including the dependence of their parameters on graphene conductivity. 
We have also predicted the existence of nonlinear surface modes in the form of discrete surface solitons.

\section*{Acknowledgements}

This work was partially supported by the European Regional Development Fund (ERDF) through the COMPETE program, the Australian National University, and the Portuguese Foundation for Science and Technology (FCT) through Grant No. PEst-C/FIS/UI0607/2013. We acknowledge support from the EC under the Graphene Flagship (Contract No. CNECT-ICT-604391). The authors thank I. Iorsh and I. Shadrivov for useful discussions and suggestions.

\begin{appendix}
\section{Nonlinear current in graphene under trigonal warping}
\label{sec:appendix}
\begin{figure}
\includegraphics[width=8.5cm]{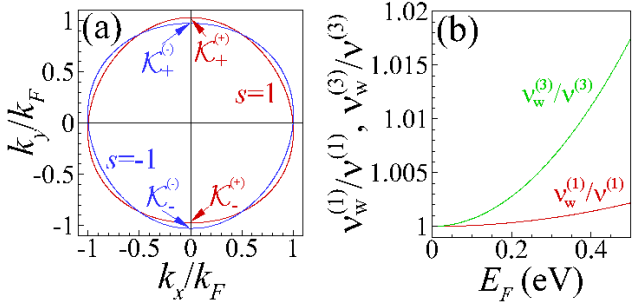} 
\protect\caption{(Color online) (a) 2D Fermi surface (for Fermi energy $E_F=\epsilon^{(s)}(\vec{k})=0.5\,$eV) of graphene with trigonal warping (\ref{eq:warping}) near two Dirac points: $s=1$ (red line) or $s=-1$ (blue line). (b) Ratios $\nu^{(1)}_w/\nu^{(1)}$  and $\nu^{(3)}_w/\nu^{(3)}$ as functions of the Fermi energy $E_F$.\label{fig:warping}}
\end{figure}
In the hexagonal lattice of a monolayer graphene each of the carbon atoms is connected to its three nearest neighbors through vectors
$\vec\delta_{1}=(-\frac{1}{2},\frac{\sqrt{3}}{2})a_0$,
$\vec\delta_{2}=(-\frac{1}{2},-\frac{\sqrt3}{2})a_0$,
and $\vec\delta_{3}=(1,0)a_0$, where $a_0$ is the carbon-carbon interatomic distance. The spectrum of charge carriers in graphene can be obtained by the standard procedure (see, e.g., Ref. \cite{c:Nuno-review}) and in the conduction band it is represented as
\begin{eqnarray}
\epsilon\left(\vec{q}\right)=t_0\left[3+2\cos\left(\sqrt{3}q_ya_0\right)\right.\nonumber\\
\left.+4\cos\left(\frac{\sqrt{3}}{2}q_ya_0\right)\cos\left(\frac{3}{2}q_xa_0\right)\right]^{1/2},
\end{eqnarray}
where $\vec{q}$ is the wave vector in the graphene plane and $t_0$ is the nearest-neighbor hopping energy. Notice that the first Brillouin zone, $-\frac{\pi}{3a_0}\le q_x \le \frac{\pi}{3a_0}$, $-\frac{2\pi}{\sqrt{3}a_0}\le q_y \le \frac{2\pi}{\sqrt{3}a_0}$, contains two Dirac points $\vec{K}_D^{(s)}=\left(0,s\frac{4\pi}{3\sqrt{3}a_0}\right)$ ($s=\pm 1$), in the vicinity of which the expansion $\vec{q}=\vec{K}_D^{(s)}+\vec{k}$ results in \cite{c:peeters-JPCM,c:PhysRevLett.113.156603}
\begin{eqnarray}
\epsilon^{(s)}\left(\vec{k}\right)=\hbar v_F\left[k_x^2+k_y^2 +s\frac{a_0}{2}k_y\left(3k_x^2-k_y^2\right)\right]^{1/2},
\label{eq:warping}
\end{eqnarray}
where the Fermi velocity $v_F=3t_0a_0/(2\hbar)$.

From Eq. (\ref{eq:warping}) it follows that shapes of the spectrum in the vicinity of Dirac points $\vec{K}_D^{(+)}$ and $\vec{K}_D^{(-)}$ are not equivalent, as demonstrated in Fig. \ref{fig:warping}(a). As a result, due to the trigonal warping nonequilibrium distribution functions,
\begin{eqnarray}
f^{(s)}(\vec{k},t)=\gamma e^{-\gamma t}\int_{-\infty}^{t}dt^{\prime}e^{\gamma t^{\prime}}\nonumber\\
\times\Theta\left\{E_F-\epsilon^{(s)}[k_{x}+H\left(t,t^{\prime}\right),k_{y}]\right\}\label{eq:BeqSol-s}
\end{eqnarray}
are different for valleys $s=\pm 1$ [compare with Eq. (\ref{eq:BeqSol})]. In this case the total current in the armchair ($x$) direction can be expressed as
\begin{eqnarray}
j_x=-2\frac{e}{(2\pi)^{2}}\sum_{s=\pm 1}\int d\vec{k}\, f^{(s)}(\vec{k},t)\,\frac{\partial\epsilon^{(s)}\left(\vec{k}\right)}{\hbar\partial k_x}=\nonumber\\
-\frac{ev_{F}}{2\pi^{2}}\gamma e^{-\gamma t}\int_{-\infty}^{t}dt^{\prime}e^{\gamma t^{\prime}}\sum_{s=\pm 1}I^{(s)}_w\left(t,t^{\prime}\right)\,,\label{eq:current_x-warp}
\end{eqnarray}
where
\begin{eqnarray}
  I^{(s)}_w\left(t,t^{\prime}\right)=\int_{{\cal K}_-^{(s)}}^{{\cal K}_+^{(s)}}dk_y^{\prime}\int_{-B(k_y^{\prime})}^{B(k_y^{\prime})}dk_x^{\prime}\nonumber\\ \frac{\left\{k_x^{\prime}-H\left(t,t^{\prime}\right)\right\}\left(1+s\frac{3a_0}{2}k_y^{\prime}\right)}{\sqrt{\left\{k_x^{\prime}-H\left(t,t^{\prime}\right)\right\}^2\left(1+s\frac{3a_0}{2}k_y^{\prime}\right) +k_y^{\prime2}-s\frac{a_0}{2}k_y^{\prime3}}},\label{eq:integral-warp}
\end{eqnarray}
with new variables $k_x^{\prime}=k_x+H\left(t,t^{\prime}\right)$ and $k_y^{\prime}=k_y$. In Eq. (\ref{eq:integral-warp}) the limits of integration are
$$
B(k_y^{\prime})=\sqrt{\frac{k_F^2-k_y^{\prime2}+s\frac{a_0}{2}k_y^{\prime3}}{1+s\frac{3a_0}{2}k_y^{\prime}}},
$$
and ${\cal K}^{(s)}_{\pm}$ are the roots of the equation $B({\cal K}^{(s)}_{\pm})=0$ [depicted in Fig. \ref{fig:warping}(a)]. Performing the integration with respect to $k_x^{\prime}$, and expanding the result in series up to the third order [similar to Eq. (\ref{eq:I-expand})], we obtain
\begin{eqnarray}
I^{(s)}_w\left(t,t^{\prime}\right)=-k_F\eta_1^{(s)}\left(k_F\right)H\left(t,t^{\prime}\right)\nonumber\\
+\frac{\eta_3^{(s)}\left(k_F\right)}{8k_F}H^3\left(t,t^{\prime}\right).\label{eq:I-expand-warp}
\end{eqnarray}
Here $\eta_1^{(s)}$ and $\eta_1^{(s)}$ are the following integrals
\begin{eqnarray}
\eta_1^{(s)}(k_F)=2\int_{{\cal K}_-^{(s)}/k_F}^{{\cal K}_+^{(s)}/k_F}d\kappa\sqrt{1-\kappa^2+s\frac{a_0k_F}{2}\kappa^3}\label{eq:eta1}\\
\times\sqrt{1+s\frac{3k_Fa_0}{2}\kappa},\nonumber\\
\eta_3^{(s)}(k_F)=8\int_{{\cal K}_-^{(s)}/k_F}^{{\cal K}_+^{(s)}/k_F}d\kappa\sqrt{1-\kappa^2+s\frac{a_0k_F}{2}\kappa^3}\label{eq:eta3}\\
\times\kappa^2\sqrt{1-s\frac{k_Fa_0}{2}\kappa}\left(1+s\frac{3k_Fa_0}{2}\kappa\right)^{\frac32}.\nonumber
\end{eqnarray}
Due to the fact that ${\cal K}^{(s)}_{\pm}=-{\cal K}^{(-s)}_{\mp}$, the integrals (\ref{eq:eta1}) and (\ref{eq:eta3}) have the properties $\eta_1^{(+)}(k_F)=\eta_1^{(-)}(k_F)$ and $\eta_3^{(+)}(k_F)=\eta_3^{(-)}(k_F)$.

Further, in full similarity with Sec. \ref{sec:current}, substituting the expansion (\ref{eq:I-expand-warp}) into Eq. (\ref{eq:current_x-warp}), integrating with respect to $t^{\prime}$, and then putting $\gamma=0$, we obtain a final expression for the nonlinear current in the form
\begin{equation}
j_{x}=i\left[\nu^{(1)}_w-\nu^{(3)}_w\left|E_{0}\right|^{2}\right]E_{0}\exp(-i\omega t),
\label{eq:current_short_2}
\end{equation}
where
\[
\nu^{(1)}_w=\sigma_{0}\frac{4E_{F}\eta_1^{(+)}\left(k_F\right)}{\pi^2\hbar\omega},\qquad\nu^{(3)}_w=\sigma_{0}\frac{9e^{2}v_{F}^{2}\eta_3^{(+)}\left(k_F\right)}{2\pi^2 E_{F}\hbar\omega^{3}}.
\]

Thus, when trigonal warping is taken into account, for finite $E_F$ both linear $\nu_w^{(1)}$ and nonlinear $\nu_w^{(3)}$ parts of the conductivity slightly exceed the values $\nu^{(1)}$ and $\nu^{(3)}$ calculated within the  Dirac cone approximation [see Fig.\ref{fig:warping}(b)] and the  ratios $\nu_w^{(1)}/\nu^{(1)}$ and $\nu_w^{(3)}/\nu^{(3)}$ grow monotonically with an increase of $E_F$. Nevertheless, taking into account the trigonal warping gives only relatively small correction to the conductivity: for typical values of the Fermi energy in graphene, $E_F\lesssim 0.5\,$eV, the difference between $\nu_w^{(1)}$ and $\nu^{(1)}$ is within 0.3$\,\%$, while that between the nonlinear conductivities $\nu_w^{(3)}$ and $\nu^{(3)}$ is below 2$\,\%$.
\end{appendix}

\end{document}